\documentclass[preprint]{aastex}

\shorttitle{GL 2591: probing the birth of a hot core?}
\shortauthors{Boonman et al.}

\input psfig
\usepackage{graphics}
\def\lteq{{_<\atop{^-}}}
\def\gtsim{{_>\atop{^\sim}}}
\def\ltsim{{_<\atop{^\sim}}}

\begin{document}

\title{Highly abundant HCN in the inner hot envelope of GL 2591: probing the 
birth of a hot core?}

\author{A.M.S. Boonman}
\affil{Sterrewacht Leiden, P.O. Box 9513, 2300 RA Leiden, 
The Netherlands}

\author{R. Stark and F.F.S. van der Tak}
\affil{Max-Planck-Institut f\"ur Radioastronomie, Auf dem H\"ugel 69, 
D-53121 Bonn, Germany}

\author{E.F. van Dishoeck}
\affil{Sterrewacht Leiden, P.O. Box 9513, 2300 RA Leiden, 
The Netherlands}

\author{P.B. van der Wal and F. Sch\"afer}
\affil{Max-Planck-Institut f\"ur Radioastronomie, Auf dem H\"ugel 69, 
D-53121 Bonn, Germany}

\and

\author{G. de Lange and W.M. Laauwen}
\affil{Space Research Organisation of the Netherlands (SRON), 
P.O. Box 800, 9700 AV Groningen, The Netherlands}

\begin{abstract}

We present observations of the $\nu_2$=0 and vibrationally excited
$\nu_2$=1 $J$=9--8 rotational lines of HCN at 797~GHz toward the
deeply embedded massive young stellar object GL~2591, which provide
the missing link between the extended envelope traced by lower-$J$
line emission and the small region of hot ($T_{\rm ex} \ge 300$~K),
abundant HCN seen in 14~$\mu$m absorption with the {\it Infrared Space
Observatory (ISO)}. The line ratio yields $T_{\rm ex}$=
720$^{+135}_{-100}$~K and the line profiles reveal that the hot gas
seen with {\it ISO} is at the velocity of the protostar, arguing against a
location in the outflow or in shocks. Radiative transfer calculations
using a depth-dependent density and temperature structure show that
the data rule out a constant abundance throughout the envelope, but
that a model with a jump of the abundance in the inner part by two
orders of magnitude matches the observations. Such a jump is
consistent with the sharp increase in HCN abundance at temperatures
$\gtsim$230~K predicted by recent chemical models in which atomic
oxygen is driven into water at these temperatures. 
Together with the evidence for ice evaporation in this source, this result 
suggests that we may be witnessing the birth of a hot core. Thus, GL~2591 may
represent a rare class of objects at an evolutionary stage just preceding
the `hot core' stage of massive star formation.

\end{abstract}


\section{Introduction}

Molecules are important tools to investigate the physical and chemical
structure of the envelopes around 
massive young stars. Physical models are a prerequisite
for determining molecular abundance profiles, which in turn are powerful
evolutionary indicators (van Dishoeck \& Blake 1998).
HCN is a particularly important molecule 
because it has pure rotational lines in the submillimeter and ro-vibrational
lines in the infrared part of the spectrum, 
which can both be observed from the ground. The combination of
submillimeter emission and infrared absorption gives strong
constraints on the structure and geometry of the 
sources (Carr et al. 1995; van der Tak et
al. 1999, 2000). It also provides independent information on the
level populations and thus allows a study of the excitation mechanisms
present in these objects.  HCN is one of the more abundant
nitrogen-bearing species in dense clouds and plays a key role in the
nitrogen chemistry in hot cores (Viti \& Williams 1999; Rodgers \& Charnley 
2001). Therefore it might also be a good candidate to probe the on-set of the 
`hot core' phase in massive star-forming regions. This is the phase just 
before 
the ultra compact \ion{H}{2} region is formed, in which the massive embedded 
young star still has a high accretion rate, but begins to 
ionize its surrounding material, with thermal pressure creating a warm dense 
neutral shell. The hot core phase is 
characterized by weak radio continuum emission, evaporation of ices, and 
high-temperature chemistry ($T\gtsim$300 K) leading to 
enhanced gas-phase molecular abundances with respect to earlier evolutionary 
phases and the production of complex molecules, such as CH$_3$OCH$_3$ and 
CH$_3$CN (Kurtz et al. 2000).\

HCN has been detected toward many massive star-forming regions in
the submillimeter and the infrared (e.g. Ziurys \& Turner 1986; 
Evans, Lacy, \& Carr 1991; Carr et al. 1995). 
Typical HCN abundances toward massive protostars 
that do not show the presence of a hot core are 
$\sim$10$^{-9}$--10$^{-8}$, derived from submillimeter 
observations (Schreyer et al. 1997).
Even most hot cores, which are predicted to have enhanced gas-phase 
HCN abundances, show typical HCN 
abundances of $\sim$10$^{-9}$--10$^{-8}$, based on low-$J$ submillimeter lines 
(Hatchell, Millar, \& Rodgers 1998). On the other hand, 
observations with the {\it Infrared Space Observatory (ISO)} of the $\nu_2$ 
ro-vibrational band around 14 $\mu$m toward a dozen massive young stars 
thought to precede the `hot core' phase have revealed strong HCN absorption 
indicating that hot gas ($T_{\rm ex} >$ 300~K) is present 
with abundances up to $10^{-6}$ (Lahuis \& van Dishoeck 2000). 
Such high HCN abundances have been derived in very few massive star-forming 
regions from millimeter data, and, except for Orion, the high estimates are 
mostly based on interferometric observations of a single 
low-$J$ isotopic transition, leading to large uncertainties 
in the derived HCN abundances (Carral \& Welch 1992).
Most of the well-known hot cores such as W~3(H$_2$O) and G~34.3, which show 
a wealth of complex organic molecules, are too weak at mid-infrared 
wavelengths to observe through infrared absorption with 
current instrumentation. The high frequency submillimeter data presented 
here provide an alternative method to probe abundant HCN.\ 

The low spectral resolution of the {\it ISO} data 
($R=\lambda/\Delta \lambda \sim$1500
at 14~$\mu$m) allows no kinematical information to be derived for the
hot abundant HCN gas and therefore its origin is not clear, 
in particular whether it is located in the inner hot part of the envelope or 
produced in shocks. Most of the sources
showing hot HCN gas also possess significant outflows, so that shock
chemistry cannot be ruled out. This is especially the case for the
massive protostar GL~2591, where a significant outflow component is
seen in infrared CO absorption lines (Mitchell et al. 1989; van
der Tak et al. 1999). High-resolution heterodyne spectroscopy of
high-$J$ rotational lines can distinguish between these two
explanations. We present here observations of the HCN $\nu_2$=0 $J$=9--8
line and the first detection of the vibrationally-excited
$\nu_2$=1 $J$=9--8 line of HCN at 797.330 GHz toward the massive
protostar GL~2591. So-far, the highest observed rotational transitions
of HCN in this source are the $\nu_2$=0 $J$=4--3 and the vibrationally excited
$\nu_2$=1 $J$=4--3 line (van der Tak et al. 1999). The advantage of the
$J$=9--8 transitions is that they have higher excitation energies and
can be observed at higher angular resolution than the low-$J$
transitions. In this way, the warmer inner part of the molecular
envelope is uniquely probed. Using the low-$J$ HCN, H$^{13}$CN, and HC$^{15}$N 
rotational lines of van der Tak et al. (1999), only 
an HCN abundance for the outer envelope of GL~2591 could be derived. The 
addition of the $J$=9--8 lines allows us for the first time to construct 
an HCN abundance profile for both the inner and outer envelope. 
This will help to determine the evolutionary state of this massive 
protostar, especially whether it has already started the formation of a 
hot core.

\section{Observations}

The observations of the HCN $\nu_2$=0 and $\nu_2$=1 $J$=9--8 lines 
were made with the MPIfR/SRON 800 GHz heterodyne
spectrometer at the James Clerk Maxwell Telescope (JCMT)\footnote{The 
James Clerk Maxwell Telescope is operated by the Joint Astronomy Centre, 
on behalf of the Particle Physics and Astronomy Research Council of the 
United Kingdom, the Netherlands Organization for Scientific Research, and 
the National Research Council of Canada.} on 2000 April 19.
The spectrometer is based on the MPIfR 795--880 GHz 
quasioptical 
SIS-receiver employing standard niobium junction technology with planar 
tuning circuits fabricated from normal conducting aluminium. The receiver 
uses an InP Gunn oscillator followed by a doubler and a tripler stage. 
Details of the setup are described in Sch\"afer et al. (1997ab). For the 
measurements described here a prototype waveguide 
mixer with a diagonal horn was used which consists of a fixed-tuned waveguide 
mixer with a Nb SIS junction with NbTiN and Al wiring layers. These devices 
were fabricated at the University of Groningen. Details on the fabrication of 
similar SIS devices can be found in Jackson et al. (2000). This resulted in 
a double sideband receiver noise temperature of about 550 K within a band 
of 50 GHz centered at 810 GHz. The receiver has an intermediate frequency 
of 3.5 GHz and a bandwidth of 1 GHz. 
Details of the receiver will be described 
elsewhere (Stark et al., in preparation). The double sideband observations 
were taken under dry 
weather conditions yielding single sideband system temperatures of about 
6000 K. The beam size at this frequency is about 8 arcsec full width 
at half maximum (FWHM) and the main beam efficiency is 
$\eta_{\rm MB}\sim$0.2. The absolute calibration uncertainty is estimated 
at 50\%. The digital autocorrelator spectrometer backend of the JCMT was used 
with a bandwidth of 500 MHz, yielding a spectral resolution of 378 kHz 
($\sim 0.14$ km s$^{-1}$). 
The observed spectrum of the massive protostar GL~2591 
($\alpha$(1950)=20$^h$ 27$^m$ 35.93$^s$, $\delta$(1950)=+40$^{\circ}$ 
01$^\prime$ 14$^{\prime\prime}$.9), taken by beam switching over 
180$^{\prime\prime}$, is shown in 
Fig. 1 and the integrated line intensities are listed in Table 1.

\section{Results}

Two lines are detected in the spectrum of GL~2591 (Fig. 1). The 
strongest is identified as the $J$=9--8 rotational transition of HCN in the 
vibrational ground state $\nu_2$=0. 
This is the first detection of this line outside Orion 
(Stutzki et al. 1988). 
The other line is the first detection of the vibrationally excited $\nu_2$=1 
$J$=9--8 line of HCN at 797.330 GHz in a massive protostar.
This line arises 
from $\sim$1200 K above the vibrational ground state and is therefore an even 
better probe of the warm gas in this protostar 
than the $\nu_2$=0 $J$=9--8 transition which has an excitation 
energy of $\sim$190~K.
Both lines are resolved and can be fit by a single gaussian.
The width of the $\nu_2$=0 $J$=9--8 line is 5.8$\pm$0.4 
km s$^{-1}$ and of the $\nu_2$=1 line 4.4$\pm$0.4 km s$^{-1}$. 
The position of the line center of V$_{\rm LSR}$=--5.1$\pm 0.5$ 
km s$^{-1}$ for the $\nu_2$=0 and of V$_{\rm LSR}$=--6.1$\pm 0.5$ km 
s$^{-1}$ for the $\nu_2$=1 line 
is consistent with the velocity of the quiescent envelope gas of 
V$_{\rm LSR}$=--5.5$\pm 0.2$ km s$^{-1}$ (van der Tak et al. 1999). 
Although our observation of the CO $J$=7--6 line in the 
same beam clearly shows the presence of an outflow component (Fig. 1), 
the HCN lines have no wings at a level larger than $\sim$17\% of 
the peak intensity, much lower than the $\sim$30\% level of the CO 
$J$=7--6 wings. This suggests that both HCN $J$=9--8 lines 
originate in the warm quiescent gas of the molecular envelope. 
Moreover, since the vibrationally-excited $\nu_2$=1 $J$=9--8 and the 
ground-state $\nu_2$=0 $J$=9--8 HCN line are observed in the same spectrum 
and in the same side band, their relative calibration is better than 25\%.  
Also, modeling shows that both lines are optically thin (Table 1), so that 
their relative 
strength provides a very accurate handle on the 
vibrational temperature. We derive $T_{\rm ex}$=720$^{+135}_{-100}$ K, in 
excellent agreement with the value $T_{\rm ex}$=600$^{+75}_{-50}$~K, inferred 
from the mid-infrared ro-vibrational absorption band 
observed with {\it ISO} (Lahuis \& van Dishoeck 2000). This strengthens our 
main argument that the HCN $J$=9--8 lines and the ro-vibrational absorption 
band seen with {\it ISO} probe the same hot region.
The continuum present in the spectra is due to warm dust and its level
agrees within 30\% with the 350 $\mu$m continuum 
photometry of this source (van der Tak et al. 2000).

\section{Models and analysis}

The data were analyzed using the temperature and density gradients 
derived for GL~2591 by van der Tak et al. (2000), shown in Fig.~2.
This model is based on submillimeter dust continuum and CS and H$_2$CO
line emission data, and also reproduces infrared absorption
measurements of CO. The radiative transfer and excitation of HCN
was calculated with the Monte Carlo code of Hogerheijde \& van der Tak
(2000) on a grid of 40 concentric shells, assuming spherical geometry.
The calculations include energy levels up to $J$=21 in both the
$\nu_2$=0 and $\nu_2$=1 states, and use collisional rate coefficients
by S.~Green (see http://www.giss.nasa.gov/data/mcrates). 
For transitions between 
vibrational levels, a rate coefficient of $10^{-12}$~cm$^{3}$s$^{-1}$
was assumed. Radiative excitation through the 14~$\mu$m band due to
warm dust mixed with the gas was also included, using grain opacities from 
Ossenkopf \& Henning (1994) and assuming $T_{\rm dust}$=$T_{\rm gas}$. 
No external radiation field apart from the 2.73~K cosmic background 
radiation was applied. Comparison with the observed 
emission proceeds by convolving the calculated sky brightness with the
appropriate beam pattern.\

In addition to the HCN $J$=9--8 lines discussed here, 
the low-$J$ HCN and isotopic lines from van der Tak et al. (1999) 
(see Table 1) are used for comparison with the model. 
It was found that this model cannot reproduce the new data using
the HCN abundances $x$(HCN)$=n$(HCN)/$n$(H$_2$) of a few times 10$^{-8}$
indicated by the low-$J$ isotopic lines. This
point is illustrated by Model~1 in Table~1. However, using the
abundance of $\sim 10^{-6}$ indicated by the $\nu_2$=1 lines and the {\it ISO} 
observations overproduces the low-$J$ line data by an order of magnitude.
Therefore, a model has been used with different HCN abundances in the
inner and outer parts of the circumstellar envelope (Fig.~3). The
`jump' is applied at $\approx 350$~AU from the star where $T$=230~K.
Chemical models predict that at this gas temperature,
reactions of O and OH with H$_2$ drive atomic oxygen into water 
(Charnley 1997; Rodgers \& Charnley 2001). 
Interferometric observations of HCN $J$=1--0 already indicated that
the HCN enhancement must be limited to a region smaller than $1500$~AU
where $T>120$~K and is therefore due to gas-phase reactions rather
than ice evaporation at $T\sim 90$~K (van der Tak et al.\ 1999).\ 

Increasing the HCN abundance in the inner envelope by a factor of 10
(Model 2) still underproduces the observed integrated intensities
(Table 1). An HCN abundance of $x$(HCN)=10$^{-6}$ in the inner
envelope, combined with the outer envelope abundance of
$x$(HCN)=10$^{-8}$ (Model 3) reproduces the observed values of
the HCN lines within $\sim$10\% (Table 1). The inner envelope HCN
abundance $x$(HCN)=10$^{-6}$ corresponds well with the value
$x$(HCN)=(6.6$\pm$1.0)$\times$10$^{-7}$ derived for the hot gas from
{\it ISO} observations. The HCN column density derived from
the latter model (Table 1) also matches best the value of 
$N$(HCN)=(4$\pm$0.6)$\times$10$^{16}$ cm$^{-2}$ derived from the
same observations by Lahuis \& van Dishoeck (2000) and the value for
the warm ($T\ge$200~K) gas of $N$(HCN)$\sim$(3--4)$\times$10$^{16}$
cm$^{-2}$ derived from ground-based observations by Carr et
al. (1995). The H$^{13}$CN and HC$^{15}$N lines are reproduced within
a factor of about 2.\

Decreasing the inner radius of the model (Fig. 2) by a factor of 3 
and extrapolating the temperature and density profile yields similar 
(within 10\%) results for Model 3. The corresponding HCN column density 
of $N$(HCN)=6.6$\times$10$^{16}$ cm$^{-2}$ is somewhat larger than that found 
from the infrared observations.\
  
Our analysis shows that both $\nu_2$=1 lines and the $\nu_2$=0 $J$=9--8 line 
are very sensitive to different abundances in the inner envelope,
contrary to the $\nu_2$=0 $J$=4--3 line which is optically thick (Table 1)
and traces only the outer envelope.
The low-$J$ isotopic lines are also sensitive to different abundances
in the inner envelope, suggesting that the warm gas contributes
significantly to their observed line profiles. This is consistent with
the maximum optical depths of $\lteq$0.33 derived for these lines (Table 1).

\section{Discussion}

The HCN abundance of $\sim$10$^{-6}$ found here for the inner 
($\ltsim 350$ AU) hot envelope of GL~2591 is much higher than predicted 
by gas-phase models of cold dense clouds of $\sim$10$^{-9}$--10$^{-8}$ 
(Lee, Bettens, \& Herbst 1996; Millar, Farquhar, \& Willacy 1997; Herbst, 
Terzieva, \& Talbi 2000). 
Models for high-temperature regions predict, however, HCN abundances of up to 
$\sim$10$^{-6}$--10$^{-5}$ (Caselli, Hasegawa, \& Herbst 1993; Charnley 1997;
Viti \& Williams 1999; Rodgers \& Charnley 2001). 
While the present data do not allow a
precise determination of where the jump occurs, our finding is
consistent with the recent chemical models by Charnley (1997) and 
Rodgers \& Charnley (2001), who show that
above a critical temperature of $230-300$~K, the reactions
O + H$_2$ $\rightarrow$ OH + H; OH + H$_2$ $\rightarrow$ H$_2$O + H 
drive most of the atomic oxygen into H$_2$O. 
This results in low gas-phase O$_2$ 
and enhanced atomic C abundances, since gas-phase O$_2$ is one of the 
principal destroyers of atomic C. The formation of atomic C is through 
cosmic-ray induced photodissociation of CO and thus is not changed. 
The formation of N$_2$ through NO + N $\rightarrow$ N$_2$ + O is also 
suppressed at high temperatures, since 
less NO is available. This provides more reactive atomic nitrogen in the 
gas-phase. These enhanced atomic C and N abundances result in 
a significantly increased gas-phase HCN abundance at temperatures 
$\gtsim$230--300~K. 
Indeed, a recent chemical model by S. Doty (2001, private communication) 
using the temperature and density gradient of GL~2591 predicts such a sharply 
increasing HCN abundance profile.
The prediction by these models of a large fraction of oxygen 
in atomic form in the outer envelope and in water in the inner part 
is consistent with the non-detection of molecular oxygen by the 
Submillimeter Wave Astronomy Satellite in this region 
(E. Bergin 2000, private communication). 
HCN ice has not been detected yet and,
as discussed above, interferometric observations  
indicate that the HCN enhancement cannot be explained by 
ice evaporation alone. However, ice evaporation 
of other species such as H$_2$O, CO$_2$, 
and C$_2$H$_2$ has occurred for GL~2591 (van Dishoeck 1998). Together
with the evidence for high-temperature chemistry inferred for HCN,
this suggests that we may be witnessing the formation of a hot
core in GL~2591. Thus, GL~2591 may
represent a rare class of objects at an evolutionary stage just preceding
the `hot core' stage of massive star formation.

Inclusion of a chemistry network in the modeling will allow a
refinement of our abundance profile. Interferometric observations of 
high-$J$ HCN transitions at sub-arcsecond resolution are needed to
determine the temperature at which the jump takes place more 
accurately. 
So-far, such a high HCN abundance as derived for the inner part of the 
molecular envelope in GL~2591 has been observed in very few massive 
star-forming regions other than Orion. 
Therefore observations of high-$J$ HCN vibrational ground-state and 
vibrationally-excited $\nu_2$=1 rotational lines toward other massive 
protostars, including hot cores, will allow a comparison of abundance profiles
between these sources and search for evolutionary effects.
Finally, high-resolution observations of other `hot core' molecules, 
such as CH$_3$OCH$_3$ in the submillimeter or C$_2$H$_2$ in the infrared, 
may confirm the presence of a hot core in GL~2591.

\acknowledgments

The support of Rolf G\"usten, Ian Robson, and Paul Wesselius in bringing
the MPIfR/SRON 800 GHz heterodyne instrument to the JCMT is greatly
appreciated. It is a pleasure to thank the JCMT staff and the MPIfR 
division for submillimeter technology for their outstanding support. 
This work was partly supported by N.W.O. Grant 614-41-003. 
EvD is grateful to the Miller Research Institute and the Department of
Astronomy at the University of California at Berkeley for their hospitality.

\clearpage

\begin{table}
\begin{center}
\caption[]{Comparison of observed integrated intensities with
different model results}
\begin{tabular}{lllrrrrc}
\noalign{\smallskip}
\tableline
Species&Band&Transition& Model 1 & Model 2 & Model 3 & Observed &Optical depth$^{\rm a}$\\
&&& K km s$^{-1}$ & K km s$^{-1}$ & K km s$^{-1}$ &  K km s$^{-1}$&\\ 
\tableline
HCN&$\nu_2$=0&$J$=4--3 & 19.38 & 20.09 & 22.95 & 24.7$^{\rm b}$&9.9\\
            &&$J$=9--8 & 14.64 & 32.52 & 57.51 & 60.2$^{\ }$&0.14\\
   &$\nu_2$=1&$J$=4--3 &  0.11 &  0.44 &  2.87 &  2.5$^{\ }$&0.006\\
            &&$J$=9--8 &  0.39 &  3.25 & 22.10 & 17.1$^{\ }$&0.03\\
H$^{13}$CN&$\nu_2$=0&$J$=3--2 & 1.93 & -- & 3.70 & 5.5$^{\ }$&0.33\\
                   &&$J$=4--3 & 1.05 & -- & 4.72 & 8.6$^{\ }$&0.11\\
HC$^{15}$N&$\nu_2$=0&$J$=3--2 & 0.45 & -- & 1.03 & 2.4$^{\ }$&0.07\\
                   &&$J$=4--3 & 0.24 & -- & 1.41 & 3.1$^{\ }$&0.03\\
\tableline
\tableline
$N$(HCN)& \ (10$^{16}$&cm$^{-2}$)& 0.12 & 0.36 & 2.7 &\\
\tableline
\end{tabular}
\tablenotetext{\rm a}{The optical depth corresponding to Model 3}
\tablenotetext{\rm b}{The narrow component at V$_{\rm{LSR}}$=--5.7 km s$^{-1}$ 
(van der Tak et al. 1999)}
\tablecomments{Model 1: Constant abundance $x$(HCN)=10$^{-8}$\\
\hspace*{1.8cm}Model 2: $x$(HCN)=10$^{-8}$ for $T<$230 K, $x$(HCN)=10$^{-7}$ 
for $T>$230~K\\
\hspace*{1.8cm}Model 3: $x$(HCN)=10$^{-8}$ for $T<$230 K, $x$(HCN)=10$^{-6}$ 
for $T>$230~K\\
\hspace*{1.8cm}The model H$^{13}$CN and HC$^{15}$N abundances correspond 
to the HCN abundance divided by the \hspace*{1.8cm}isotopic ratios 
$^{12}$C/$^{13}$C=60 and $^{14}$N/$^{15}$N=270 respectively 
(Wilson \& Rood 1994)}
\tablecomments{Calibration uncertainties in the observed integrated 
intensities are $\sim$50\% for the $J$=9--8 lines \hspace*{1.8cm}and 
$\sim$30\% for the 
other lines (van der Tak et al. 1999).} 
\end{center}
\end{table}

\clearpage

\begin{figure}
\begin{center}
\psfig{figure=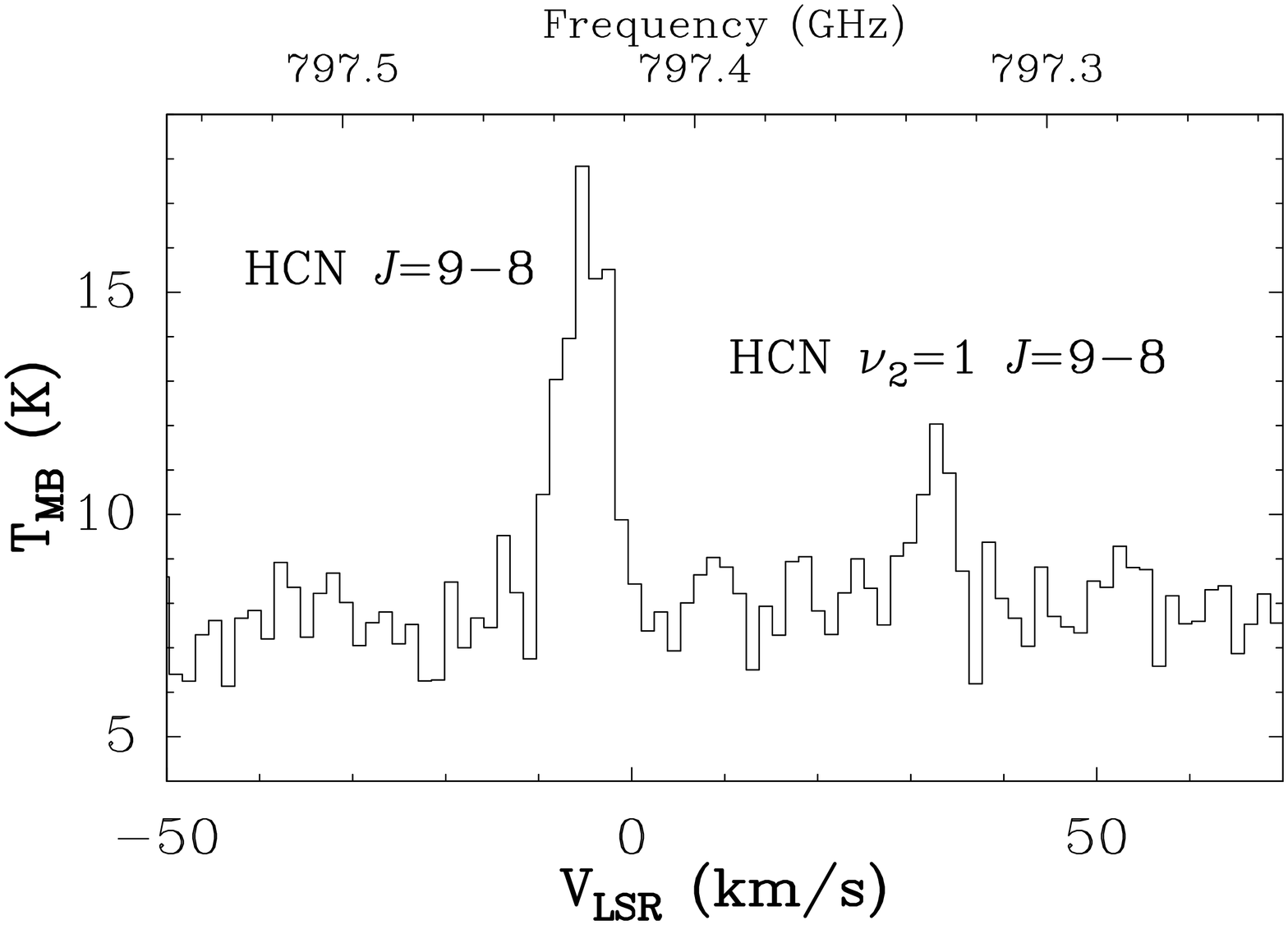,angle=270,width=10 cm}
\psfig{figure=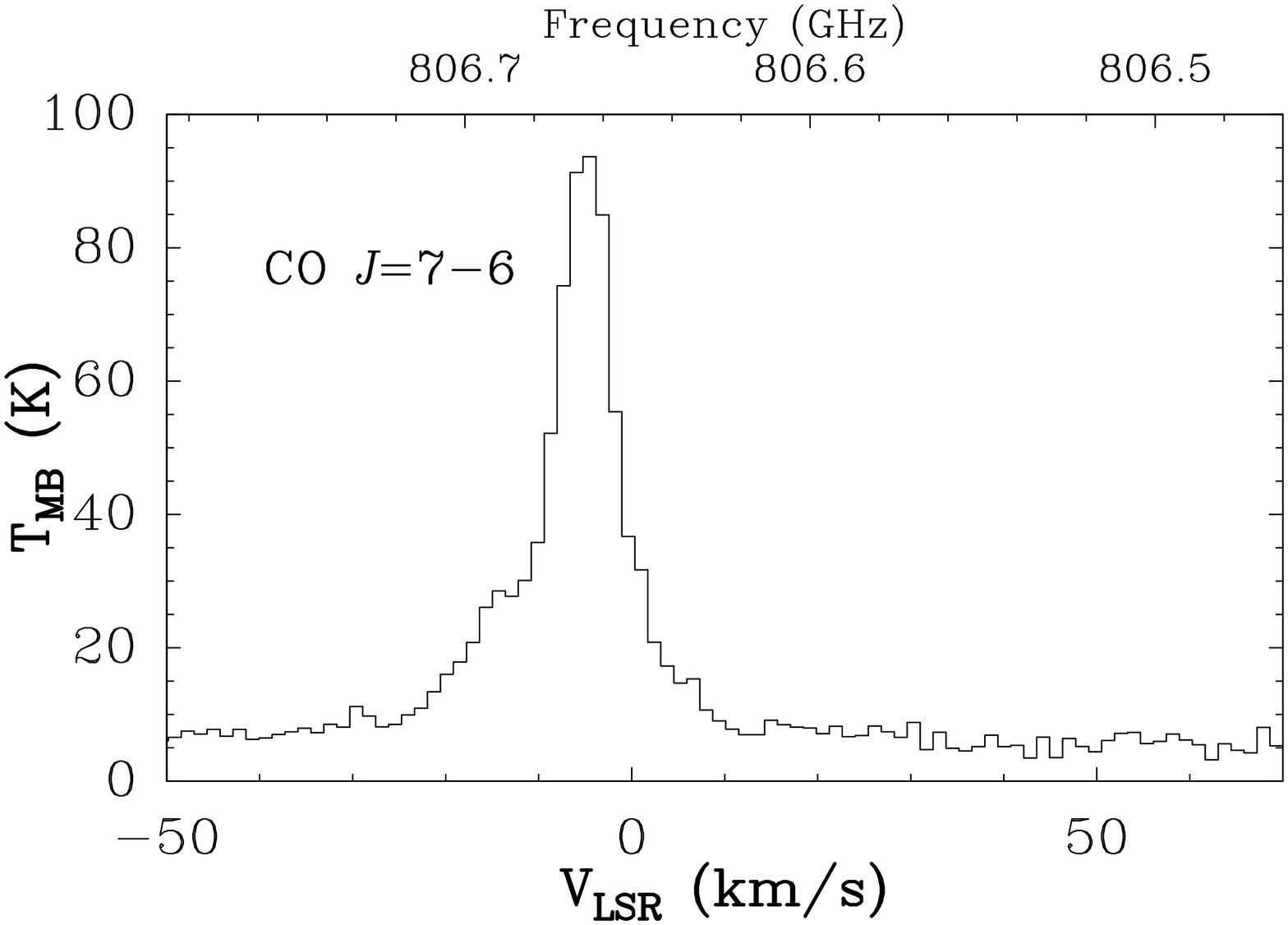,angle=270,width=10 cm}
\figcaption{Spectrum of the $\nu_2$=0 $J$=9--8 and the 
vibrationally excited $\nu_2$=1 $J$=9--8 line of HCN at 797.434 GHz and 
797.330 GHz, respectively, toward the 
massive protostar GL~2591 ({\it Top panel}). The spectrum has been smoothed 
to a spectral resolution of $\sim$1.4 km s$^{-1}$. The CO $J$=7--6 line 
at 806.652 GHz is shown
for comparison ({\it Bottom panel}). The latter spectrum shows clearly the 
presence of a wing component whereas 
the HCN $J$=9--8 lines do not show such a component. The vertical offset 
shows the submillimeter continuum due to warm dust, which 
agrees within 30\%  
with the 350 $\mu$m continuum 
photometry of this source (van der Tak et al. 2000).}
\end{center}
\end{figure}
\begin{figure}
\begin{center}
\psfig{figure=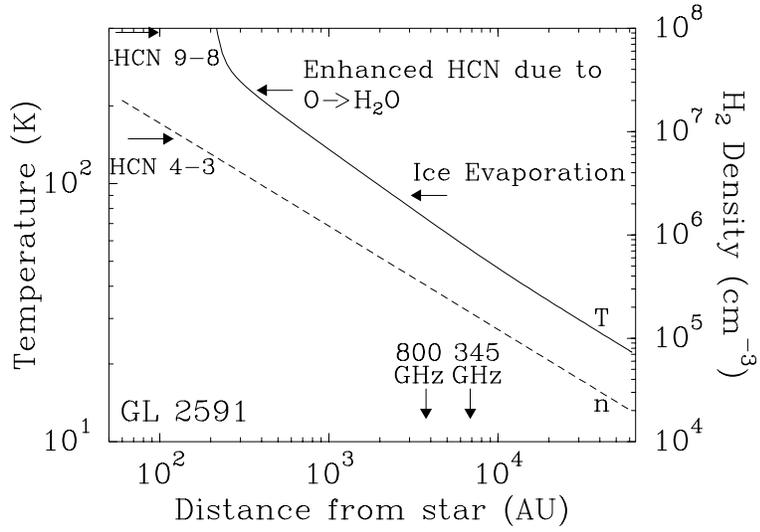,angle=270,width=11 cm}
\figcaption{Temperature and density structure of GL~2591. The critical 
densities for the $\nu_2$=0 $J$=9--8 and $\nu_2$=0 $J$=4--3 lines are 
indicated as well as the JCMT beam sizes at the different frequencies (adapted 
from van Dishoeck \& van der Tak 2000).}
\end{center}
\end{figure}
\begin{figure}
\begin{center}
\psfig{figure=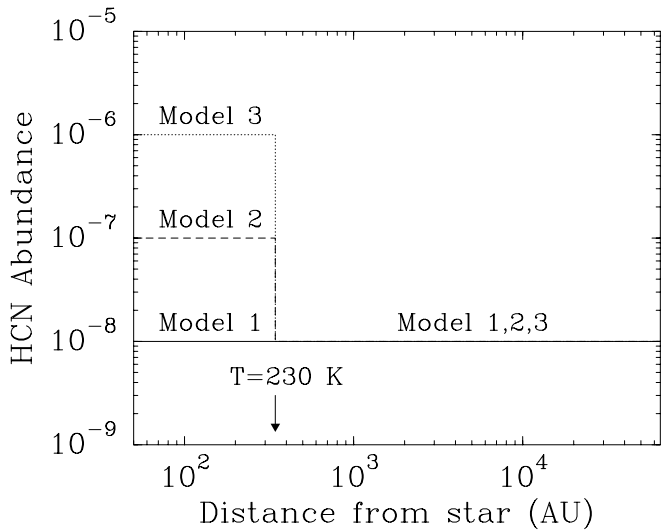,angle=270,width=11 cm}
\figcaption{Different HCN abundance profiles used in the analysis of \S 4 
(see also Table 1).}
\end{center}
\end{figure}

\clearpage

\end{document}